\documentclass[10pt,a4paper]{article}

\title{\"{U}ber das Gravitationsfeld eines Massenpunktes nach der Einsteinschen Theorie}

\author{Llu\'{\i}s\ Bel\\
\emph{wtpbedil@lg.ehu.es}
}
\date{}
\begin{document}

\maketitle

\begin{abstract} 

Schwarzschild's solution of Einstein's field equations in vacuum can be written in many different forms. Unfortunately Schwarzschild's own original form is less nice looking and simple than that latter derived by Droste and Hilbert. We prove here that we can have both: a nice looking simple form and the meaning that Schwarzschild wanted to give to his solution, i.e., that of describing the gravitational field of a massive point particle.   

\end{abstract}

\section*{Introduction}

In 2003 the editorial board of {\it General Relativity and Gravitation} decided to reprint Schwarzschild's original paper \cite{Schwarzschild} as a Golden Oldie. To that end the Journal used an S. Antoci and A. Loinger's translation\,\footnote{Also posted in arXiv:Physics/9905030} from the German to English as an Appendix to an Editors note by S. Antoci and D.-E. Liebscher \cite{Antoci} which included some unconventional comments on how badly the Relativity community has misunderstood, ignored or misquoted Schwarzschild's work.

Apparently the current Editorial board of {\it General Relativity and Gravitation} felt unhappy with the Editor's note published three years before, and decided to publish Corrections to the editorial note, something that has been done by J. M. Senovilla in \cite{Senovilla}. This unusual rectification of the Editorial board is rather worrisome and it would be reassuring to know that their magazine is not committed to a single acceptable point of view on how to understand General Relativity.

Senovilla's paper is a nice peace of information about the modern way of understanding some aspects of Einstein's theory of gravitation that are centered on new concepts like those of event horizon, essential singularity or maximal extension. All this is nice geometry in the making but the point is that none of this is as yet necessary to understand that Schwarzschild's original work is a better piece of physics than the extravaganzas to which one is led with some of the extensions of Schwarzschild's solution. At least this is what we believe, hopefully among others.

We remind in the first section of this paper how and why Schwarzschild was misled to choose a complicated path to obtain his solution, the result being that his paper is significantly less simple to read than latter derivations of his solution. The following sections are meant to be an improvement of Schwarzschild's paper while remaining faithful to his endeavor to obtain a satisfactory model of the gravitational field of a point particle.

This paper contains very little new material and it has mainly a pedagogical purpose.

\section{Schwarzschild's form of Schwarzschild's solution}

Schwarzschild discovered his solution at a time when Einstein's was still preferring a non fully covariant theory. In fact he was simply privileging a coordinate condition. For this reason  he was forced to follow what it looks to us today as a quite unnatural path. Namely: to start with a system of coordinates such that the determinant of the metric be $-1$. 

What we call here Schwarzschild's form of Schwarzschild's solution is a line-element that he did not write explicitly, but that he manifestly used. That is to say\,\footnote{We shall use units such that $c=G=1$}:

\begin{eqnarray}
\label{1.2}
&& ds^2=-\left(1-\frac{\alpha}{(r^3+\rho)^{1/3}}\right)dt^2+ \nonumber \\ 
&&\hspace{2cm}\frac{r^4}{(r^3+\rho)^{4/3}}\left(1-\frac{\alpha}{(r^3+\rho)^{1/3}}\right)^{-1}dr^2+(r^3+\rho)^{2/3}d\Omega^2
\end{eqnarray}
where:

\begin{equation}
\label{1.2.1}
d\Omega^2=d\theta^2+\sin^2\theta d\varphi^2
\end{equation}
The coordinates $r,\ \theta,\ \varphi$ are no longer the coordinates he started with and in fact this line-element is an exact vacuum solution of the fully covariant Einstein´s theory.

Using the ``auxiliary function'':

\begin{equation}
\label{1.3}
R=(r^3+\rho)^{1/3}
\end{equation}
what Schwarzschild wrote explicitly was:
\begin{equation}
\label{1.1}
ds^2=-\left(1-\frac{\alpha}{R}\right)dt^2+ \left(1-\frac{\alpha}{R}\right)^{-1}dR^2+R^2d\Omega^2, 
\end{equation}
while still referring to the origin as the point with radial coordinate $r=0$.
 
The line-element (\ref{1.2}) contains two free parameters $\alpha$ and $\rho$. He knew from Einstein's work on the motion of the perihelion of Mercury that $m$ being the mass of the source of the field he had to have $\alpha=2m$, but $\rho$ was still arbitrary. He realized that the physical problem was not unique and to make it unique he decided that the coefficient of $dr^2$ should be positive on the whole interval $r>0$ and become infinite when $r$ goes to zero, where the point particle was supposed to be. This forced him to accept the condition $R=\alpha$, or equivalently to fix the free parameter to be $\rho=\alpha^3$. Actually Schwarzschild's text justifies first this value of $\rho$ and when he simplifies his line-element (\ref{1.2}) using the auxiliary function (\ref{1.3}) the value of his free parameter had already been chosen.

A few years latter Droste \cite{Droste} and Hilbert \cite{Hilbert} re-derived a new form of Schwarzschild's solution that they wrote as:

\begin{equation}
\label{1.6}
ds^2=-\left(1-\frac{\alpha}{r}\right)dt^2+ \left(1-\frac{\alpha}{r}\right)^{-1}dr^2+r^2d\Omega^2
\end{equation}
where the coefficient $r^2$ of $d\Omega^2$ was chosen beforehand as the full covariance of the theory allowed. This new form is simpler to obtain than (\ref{1.2}) and also simpler to write down and is the form which is used overwhelmingly in text-books. Notice that it can be derived directly from Schwarzschild's form following two different, but equivalent, paths:

(i) To use the definition (\ref{1.3}) of the auxiliary function $R$ as a coordinate transformation and get rid of the spurious parameter $\rho$, or (ii) choose for simplicity $\rho=0$. 

Schwarzschild could not follow the first path because he thought he was dealing with a theory which did not allow arbitrary coordinate transformations (but in fact he had already done it when he abandoned his initial coordinates for those used in (\ref{1.2})). He could have chosen the second path but he deliberately avoided it because he understood, better than Droste and Hilbert, that to construct a satisfactory physical model requires to pay attention to two ingredients: (a) the field equations, that in Einstein's theory are fully covariant, and (b) the side conditions that need not to be and are always necessary to make the model unique. 

In the next section we shall prove that there is a preferred form of Schwarz\-schild's solution that is both simple and faithful to his endeavor.

\section{Local forms of Schwarzschild's solution}

(a) {\it Main coordinate conditions}.- Let $r,\ \theta,\ \varphi$ be a system of polar space coordinates of $R^3$ with an arbitrary origin and let $t$ be a time-coordinate of $R^1$. And let us consider the space-time manifold $V_4=V_3 \times R$, $V_3$ being the sub-manifold of $R^3$ defined by the interval of $r$: 

\begin{equation}
\label{2.1}
0<r_0<r<r_1
\end{equation}

We consider an space-time line-element of class $C^\infty$ defined on $V_4$:

\begin{equation}
\label{2.2}
ds^2=g_{\alpha\beta}(x^\rho)dx^\alpha dx^\beta
\end{equation}
where $x^0=t,\ x^1=r,\ x^2=\theta,\  x^3=\varphi$, that we assume to be time independent and spherically symmetric so that explicitly it is:

\begin{equation}
\label{2.3}
ds^2=-A(r)dt^2+B(r)dr^2+2D(r)drdt+r^2C(r)d\Omega^2.
\end{equation}
The set of transformations that while changing the coefficients of (\ref{2.3}) do not change its structure is:

\begin{equation}
\label{2.4}
t\leftarrow kt+\psi(r), \quad r\leftarrow\phi(r)
\end{equation}
where $k$ is constant and the functions $\psi$ and $\phi$ of $r$ are arbitrary. 

We want to find the general solution of Einstein's equations in vacuum:

\begin{equation}
\label{2.5}
S_{\alpha\beta}=R_{\alpha\beta}-\frac12 Rg_{\alpha\beta}=0
\end{equation}
assuming that: 

\begin{equation}
\label{2.6}
A(r)>0,\ B(r)>0,\ C(r)>0,
\end{equation}
on the interval (\ref{2.1})

As it is well-known the fact that (\ref{2.5}) are covariant equations allow to choose two coordinate conditions using (\ref{2.4}). Here we shall use the two conditions $D(r)=0$, and:

\begin{equation}
\label{2.7}
A(r)B(r)=1
\end{equation}
so that the line-element (\ref{2.3}) becomes:

\begin{equation}
\label{2.8}
ds^2=-A(r)dt^2+A(r)^{-1}dr^2+r^2C(r)d\Omega^2,
\end{equation}
The only coordinate freedom that remains after this simplification is to use the transformation:

\begin{equation}
\label{2.9}
r\leftarrow r+\lambda
\end{equation}
$\lambda$ being an arbitrary parameter.

(b) {\it Conformal metric}.- We proceed as if we did not know any local form of Schwarzschild's solution and proceed to analyze on how many parameters depends the general local solution of (\ref{2.5}) when requiring (\ref{2.7}). $\lambda$ being one of them of course.

One of the immediate benefits of condition (\ref{2.7}) is that the 3-dimensional space-metric\,\footnote{see \cite{Fock}, \cite{Bel1},\cite{Bel2},\cite {Geroch}}:

\begin{equation}
\label{2.11}
d\bar s^2=(-g_{00})\left(g_{ij}-\frac{g_{0i}g_{0j}}{g_{00}}\right)dx^idx^j,
\end{equation}
which is conformal to the quotient metric on $V_3$, becomes:

\begin{equation}
\label{2.12}
d\bar s^2=dr^2+r^2\bar C(r)d\Omega^2, \quad \bar C(r)=A(r)C(r)
\end{equation}
This meaning that if $x_3$ and $x_4$ are two points of $V_3$ on a straight line through the center then:

\begin{equation}
\label{2.13}
\int_{x_3}^{x_4}{d\bar s}=r_4-r_3
\end{equation}
This provides the intervals of $r$ with an obvious geometrical interpretation. And should $r_0$ tend to zero
then $r$ would be both the euclidean and the Riemannian distance, in the sense of (\ref{2.12}), from a point with radial coordinate $r$ to the origin. This by itself justifies the choice of the condition (\ref{2.7}). 

(c) {\it Einstein's equations}.- Explicitly Eqs. (\ref{2.5}) can be written as three equations with two unknowns $A(r)$and $\bar C(r)$, of the following form:

\begin{eqnarray}
\label{2.14}
&&\bar C^2r^2(S_{00}+A^2S_{11})+2\bar CA^2S_{22}=\nonumber \\
&& \hspace{1cm} r^2\bar C^2AA^{\prime\prime}+(r^2\bar C\bar C^\prime+2r\bar C^2)AA^\prime-r^2\bar C^2(A^\prime)^2=0
\end{eqnarray}

\begin{eqnarray}
\label{2.15}
&& 4\bar CA^2S_{22}=\nonumber \\
&& \hspace{1cm} 2r^2A^2\bar C\bar C^{\prime\prime}+4rA^2\bar C\bar C^\prime+r^2(A^\prime)^2\bar C^2-r^2A^2(\bar C^\prime)^2=0
\end{eqnarray}
and:

\begin{eqnarray}
\label{2.16}
&& 4r^2\bar C^2A^2r^2S_{11}=\nonumber \\
&& \hspace{1cm}4rA^2\bar C\bar C^\prime+r^2A^2(\bar C^\prime)^2-r^2{\bar C}^2(A^\prime)^2-4A^2\bar C(1-\bar C)=0
\end{eqnarray}
where we dropped the explicit dependence on $r$ of the unknowns and where a prime means a derivative with respect to $r$. The particular linear combination (\ref{2.14}) has been chosen to have an equation that contains $A^{\prime\prime}$ but not $\bar C^{\prime\prime}$  and such that each term contains the same number of $A$'s and the same number of $\bar C$'s.  

(d) {\it Initial conditions}.- The solutions of the couple of Eqs. (\ref{2.14})-(\ref{2.15}) depend on four initial conditions at any particular point with $r=r^*$, namely:

\begin{equation}
\label{2.17}
A(r^*),\ \bar C(r^*),A^\prime(r^*),\ \bar C^\prime(r^*) 
\end{equation}
On the other hand as a consequence of the identities:

\begin{equation}
\label{2.18}
\nabla_\alpha S^\alpha_\beta=0 
\end{equation} 
the l-h-s of Eq. (\ref{2.16}) is a first integral of the system of equations (\ref{2.14})-(\ref{2.15}) and therefore it is sufficient that it holds at one point to guarantee that it holds everywhere in the domain of existence of the latter system. Therefore (\ref{2.18}) reduces the necessary number of initial conditions to three. Equivalently we can say that the general local solutions of (\ref{2.5}) that we are discussing depend on three parameters. But this does not mean of course that for any values of the initial data or the free parameters the solution that it is obtained will satisfy the conditions (\ref{2.6}) on a pre-selected interval (\ref{2.1}).

\section{Regular global forms}

(a) {\it Classification.} Now is time to remind that Schwarzschild work intended to describe the gravitational field of a point mass and therefore to replace the manifold $V_3$ defined in the preceding section by $R^3$. This requires to consider the behavior of the solutions of (\ref{2.14}), (\ref{2.15}) and (\ref{2.16}) both when $r_0\rightarrow 0$ and $r_1\rightarrow \infty$. Let us consider first the behavior at the origin. We shall say that a local solution on $V_3$ has been extended to a global one, regular at the origin, if the unknowns $A$ and $\bar C$ can be written as power series of $r$ of the following form:

\begin{equation}
\label{3.1}
A=a_0r^s+a_1r^{s+1}+\cdots, \quad C=\bar c_0r^p+\bar c_1r^{p+1}+\cdots
\end{equation}
which are convergent on $V_3-\{x_0\}$, the center of symmetry $x_0$ being the point with radial coordinate $r=0$.

Substituting these expressions above into (\ref{2.14}) and (\ref{2.15}) we obtain the indicial equations:

\begin{equation}
\label{3.2}
s(p+1)=0, \quad s^2+p(p+2)=0
\end{equation}
so that we have to distinguish four cases:

(i)\ \ s=0,\ p=0;\ (ii)\ \ s=0,\ p=-2;\ (iii)\ \ s=1,\ p=-1, and (iv)\ \ s=-1,\ p=-1

The first case is a trivial one: it leads to the solution $A=const.$, $\bar C=const.$ 

(b) {\it Case iv}. Let us consider now the last case (iv). This means that we are assuming  that the first two terms of the series (\ref{3.2}) are:

\begin{equation}
\label{3.3}
A=\frac{a_0}{r}+a_1, \quad \bar C=\frac{\bar c_0}{r}+\bar c_1
\end{equation}
Substitution of these expressions into Eqs. (\ref{2.14}), (\ref{2.15}) and (\ref{2.16}) proves that they are exactly satisfied provided that:

\begin{equation}
\label{3.4}
a_1=\frac{a_0}{\bar c_0},\ \bar c_1=1
\end{equation}
so that:

\begin{equation}
\label{3.5}
A=\frac{a_0}{r}+\frac{a_0}{\bar c_0}, \quad \bar C=\frac{\bar c_0}{r}+1
\end{equation}
is a first non trivial regular solution. Notice however that if $a_0$ is negative the constraints (\ref{2.6}) can not be satisfied on $\bar V_3$. This is indeed what happens with the Droste-Hilbert's form of Schwarzschild's solution and therefore it should be discarded from the outset as a global solution, although it remains acceptable as a local form in restricted manifolds $V_3$.

{\it The remaining cases}. Another interest of (\ref{3.5}) is that it can be used to find the remaining forms (ii) and (iii) because using the allowed transformation (\ref{2.9}) we obtain the general form depending on the three parameters $a_0,\bar c_0$ and $\lambda$:

\begin{equation}
\label{3.6}
A=\frac{a_0}{r+\lambda}+\frac{a_0}{\bar c_0},\quad  \bar C=\frac{1}{r^2}\left(1+\frac{\bar c_0}{r+\lambda}\right)(r+\lambda)^2
\end{equation}
whose behavior at the origin when $\lambda\neq 0$ is given by:

\begin{eqnarray}
\label{3.7}
A&=&\frac{a_0}{\lambda}+\frac{a_0}{\bar c_0}-\frac{a_0r}{\lambda^2}+\frac{a_0r^2}{\lambda^3}+O(r^{-3}) \\
\bar C&=&\left(\frac{\bar c_0}{\lambda}+1\right)\frac{\lambda^2}{r^2}+\left(2\left(\frac{\bar c_0}{\lambda}+1\right)\lambda-\bar c_0\right)\frac{1}{r}+1
\end{eqnarray} 
From these we can tell what are the values of $a_0,\bar c_0$ and $\lambda$ that we have to substitute in (\ref{3.6}) to obtain solutions of type (ii) and (iii) above. We obtain the type (ii) with $\lambda \neq -\bar c_0$. And we obtain the type (iii) with $\lambda = -\bar c_0$.

{\it The source is a point} To be able to identify $x_0$ as a point in the sense of the geometry of space (\ref{2.11}) we need to have $r^2\bar C= 0$ for $r=0$. The second equation above then implies:

\begin{equation}
\label{3.7.1}
\lambda =-\bar c_0
\end{equation}
This excludes the case (ii) and leaves as unique possibility the case (iii) with its two free parameters $a_0$ and $\bar c_0$.

{\it Asymptotic conditions}. Let us consider now the asymptotic conditions, i.e, the behavior when $r_1\rightarrow\infty$. The asymptodic expansion of $A$ and $\bar C$ for any of the cases (ii) to (iv) are: 

\begin{eqnarray}
\label{3.8}
A(r_1\rightarrow\infty)&=&\frac{a_0}{\bar c_0}+\frac{a_0}{r}+O(r^{-2}) \\
\bar C(r_1\rightarrow\infty)&=&1+\frac{2\lambda+\bar c_0}{r}+\frac{\lambda(\lambda+\bar c_0)}{r^2}
\end{eqnarray}

To guarantee then the asymptotic Minkowskian behavior of the space-time geometry (\ref{2.8}) and the validity of Newtonian gravity at large distances the first of this equations requires:  

\begin{equation}
\label{3.10}
\bar c_0=a_0,\ a_0=-2m 
\end{equation}

{\it The global regular form} Finally the only case that satisfies all the equivalent physical constraints that Schwarzschild demanded to his solution is a global regular form with the following line-element:

\begin{equation}
\label{3.11}
ds^2=-\frac{r}{r+2m}dt^2+\frac{r+2m}{r}dr^2+(r+2m)^2d\Omega^2
\end{equation} 
This form was discussed as early as in 1923 by M. Brillouin from a qualitative point of view, \cite{Brillouin}, and more recently by Senovilla in \cite{Senovilla2} as an example of singular extension of Schwarzschild's solution. 

\section{Summary and concluding remarks}

\ \ \ \ (i) We have proved that the line-element (\ref{3.11}) is the unique regular form of Schwarzschild's solution in the sense of our definition that has the properties mentioned below.

(ii) The form is global, i.e. its coefficients $A,\ B=1/A$ and $\ C$ satisfy the constraints (\ref{2.6}) on the whole interval $r>0$.

(iii) For large values of $r$ the force on a test particle of unit mass at rest with respect to the source of the solution, i. e. the intrinsic curvature $\Lambda$ of the time-like Killing trajectories $t=var.$, is:

\begin{equation}
\label{4.1}
\Lambda=-\frac{A^\prime}{2A} = -\frac{m}{r(r+2m)}
\end{equation}
It behaves like $m/r^2$ as in Newtonian gravity for large values of $r$, but for small values  it behaves as $m/r$.

(iv) Assuming that the geometry of space is given by (\ref{2.12}), i. e. now:

\begin{equation}
\label{4.2}
d\bar s^2=dr^2+r(r+2m)d\Omega^2
\end{equation}
$r$ is also the true Riemannian distance from a point of coordinates $r,\ \theta,\ \varphi$ to the source of the solution. The coefficient of $d\Omega^2$ goes to zero when $r$ goes to zero and this is what we mean here by $r=0$ being a point\,\footnote{We gave in \cite{Bel1} and \cite{Bel2} an intrinsic definition}, but it does it more slowly than what is the case for an Euclidean geometry. Moreover we have that the non-zero components of the Riemann tensor of (\ref{4.2}) in the co-frame of reference $dr,\ d\theta,\ d\varphi$ are:

\begin{equation}
\label{4.3}
\bar R_{2323}=-\bar R_{3131}=-\bar R_{1212}=-\frac{m^2}{r^2(r+2m)^2}
\end{equation}
Notice that they are proportional to $(m/r)^2$ and therefore at the linear approximation they are zero and the geometry is Euclidean, but in general they are singular when $r$ tends to zero. Both properties are interesting and peculiar when compared to the behavior of the space-time metric.

\section{Acknowledgements}

A. Molina and J. M. Senovilla have both contributed to improve an earlier manuscript.

\end{document}